\documentclass[aps,prb,twocolumn,groupedaddress,showpacs,twocolun]{revtex4}
\usepackage{graphicx,color}
\usepackage{subfigure}

\begin{document}
\title{\textbf{Conventional Superconductivity properties of the ternary boron-nitride Nb$_{2}$BN}}
\author{O. V. Cigarroa$^{1}$, S. T. Renosto$^{1}$, A. J. S. Machado$^{1}$}
\affiliation{$^1$Escola de Engenharia de Lorena, Universidade de S\~{a}o Paulo, P.O. Box 116, Lorena - SP, Brasil;}
\date{\today}
\email{orlandocv@usp.br}

\begin{abstract}
Superconducting bulk properties of ternary Nb$_{2}$BN are confirmed and are described by means of magnetization, electronic transport and specific-heat measurements. BCS conventional superconductivity is found with T$_{c}=4.4$ K. Critical fields H$_{c1}$(0)= 93 Oe and H$_{c2}$(0)= 2082 Oe are extrapolated by magnetic and resistivity measurements. The specific heat data reveals $\gamma=6.3$
mJ/mol K$^{2}$ and $\beta =0.293$ mJ/mol K$^{4}$ in good agreement with the BCS Theory.

\end{abstract}

\pacs{74.20.Fg, 71.28.+d, 72.80.-r}
\keywords{Nb$_{2}$BN, superconductivity, BCS}
\maketitle

\begin{center}
\textbf{I. INTRODUCTION}
\end{center}

\begin{figure} [htp]
\includegraphics[width=4.0cm]{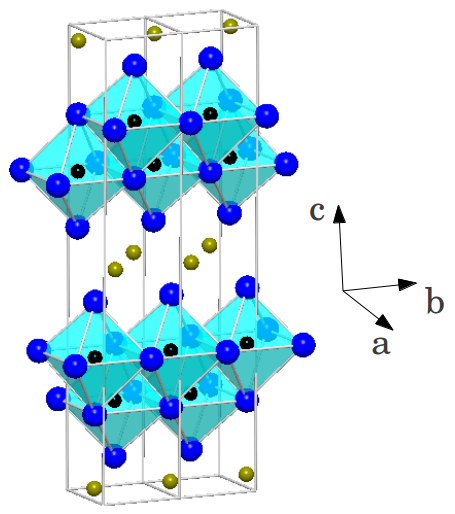}
\caption{\label{fig1} Schematics of the Mo$_{2}$BC orthorhombic (cmcm) type structure. Distorted Mo$_{6}$ octahedron layers (blue spheres) with C atoms (black spheres) on their center are separated by B zig-zag chains (green spheres) at the middle of the cell unit.}
\end{figure}

Among the known ternary carbides, Mo$_{2}$BC deserves attention due its superconducting behaviour with a T$_{c}$ ranging from 5 to 7.5 K \cite{Mo, Mo2}. Mo$_{2}$BC has an orthorhombic symmetry \cite{cel1} (no. 63, space group cmcm) with $a=3.086$, $b=17.35$ and $c=3.047$ $\AA$. Fig. \ref{fig1} illustrates such structure, where distorted Mo$_{6}$ octahedron layers are separated from each others by zig-zag B chains which pass through the trigonal prisms of Mo atoms \cite{cel2}. Carbon atoms are located at the center of the Mo octahedrons \cite{cel3, cel4}. Several efforts \cite{cel4, ef1, ef2, ef3} were made to enhance the superconducting properties of this compound by inducing chemical pressure with transition elements as M = Zr, Rh, Nb, Hf, Ta and W at the Mo sites. While T$_{c}$ of Mo$_{2-x}$M$_{x}$BC decreased with increase of $x$ for all the alloys, only the Rh-containing alloy \cite{ef3}  showed an increase up to T$_{c}=9$K. 

After the discovery of Mo$_{2}$BC  a boron-nitride with nominal Nb$_{2}$BN composition was synthesized for the first time \cite{Nb} as a thermodynamically stable compound at 1200$^{\circ}$C. In this compound Nb atoms are arranged in similar distorted octahedron and are separated by B zig-zag chains as in Mo$_{2}$BC and N atoms occupy the same position as C atoms. Lattice Parameters of Nb$_{2}$BN are quite similar to those of Mo$_{2}$BC being $a=3.17$, $b=17.85$ and $c=3.11$ $\AA$. Superconductivity with critical temperature close to 2.2 K was reported in Nb$_{2}$BN$_{x}$ (x undetermined) compound. However other properties such as specific heat or resistivity were not performed by the authors in order to confirm bulk superconductivity. In this context our results show unambiguously bulk superconductivity at 4.4 K in single phase polycrystalline samples of Nb$_{2}$BN. 

\begin{center}
\textbf{II. EXPERIMENTAL}
\end{center}

Polycrystalline Nb$_{2}$BN samples were synthesized by conventional powder solid state reaction method. High purity 300 mesh Nb (99.99) and hexagonal BN (99.999) were used. To ensure the good quality of the primary reaction BN was degassed at 1000$^{\circ}$C in vacuum for twenty four hours before its usage. Stoichiometric amounts (Nb 2:1 BN) of the reagents were weighted, mixed on an agatar mortar and pressed (4 tons) into pellets of cylindrical shape. Compressed mixtures were sealed on quartz tubes under 1 bar argon atmosphere and heated at 1200$^{\circ}$C.  Each 7 days of annealing the samples were quenched,  grounded, pressed and encapsulated again to be treated at 1200$^{\circ}$C. A complete reaction was obtained only after 28 days of annealing.              

X-ray diffraction patterns were collected with a Panalitical Empyrean X-ray diffractometer using CuK$_{\alpha}$ radiation. Rietveld refinements \cite{young} were calculated using the Fullprof suite considering a error $\chi^{2}\leq 2$ as the minimum standard. Magnetic, electric, and thermal, initial characterizations were made using a Quantum Design PPMS Evercool II. Magnetization (\textit{M}) measurements were obtained using a commercial VSM magnetometer (Quantum Design) in a \textit{DC} external field of 50 Oe in zero field cooling (ZFC) and field cooling (FC) conditions, on a temperature range (\textit{T}) from 2 to 20 K. Magnetization (M) versus applied field (H) data were acquired at constant temperatures between 2 and 5 K. Electrical resistivity measurements were performed between 1.8 and 300 K using the conventional four-point method. Thin Cu wires were welded to a regular shape sample and served as the voltage and current leads, using a high purity Ag epoxy. Applied magnetic fields were also used to estimate the upper critical field. The superconducting critical temperature (\textit{T$_{c}$}) was defined as the transition midpoint. The heat capacity measurements were determined using the relaxation method of a piece cut from the sample on a calorimetric probe coupled to the PPMS system on a temperature range from 2 to 10 K. 

\begin{center}
\textbf{III. RESULTS AND DISCUSSION}
\end{center}

\begin{figure} [htp]
\includegraphics[width=8.5cm,height=6.5cm]{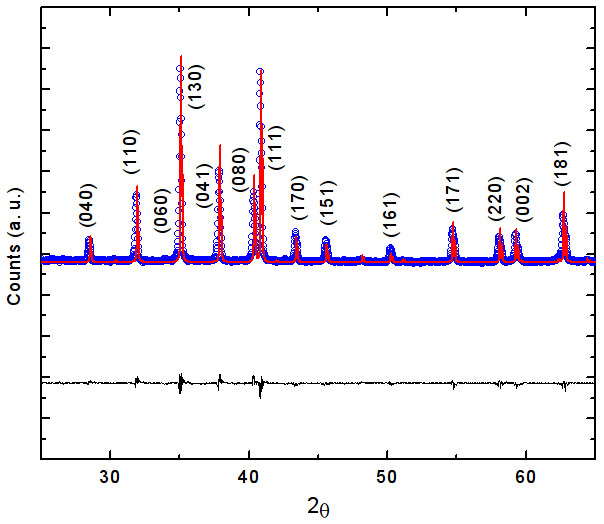}
\caption{\label{fig2} X-ray diffraction pattern for polycrystalline  Nb$_{2}$BN. Experimental pattern (blue dots) is compared with the simulated diffraction pattern (red line) and the difference between them (black line) as calculated in the Rietveld refinement.} 
\end{figure}

Figure \ref{fig2} shows the diffraction pattern of a typical  Nb$_{2}$BN sample after 28 days of heat treatment. All reflections can be indexed with the orthorhombic Mo$_{2}$BC structure with space group \textit{cmcm} and without any trace of secondary phases within the limits of this technique. Rietveld refinements led to the following occupancies: Nb-1 atoms occupy the \textit{4c} (0, 0.721, 0.25), Nb-2 atoms occupy the \textit{4c} (0, 0.3139, 0), N atoms occupy the \textit{4c} (0, 0.192, 0) and B atoms the \textit{4c} (0, 0.4731, 0.25) positions. These results are in good agreement with the ones reported earlier \cite{Nb}. The Nb-1  bonding distance have a slightly difference (2.22 and 3.01 \AA) with the original Mo-1 bonding dintance (2.11 and 3.086 \AA) in Mo$_{2}$BC.  

\begin{figure} [htp]
\includegraphics[width=8.7cm,height=6.5cm]{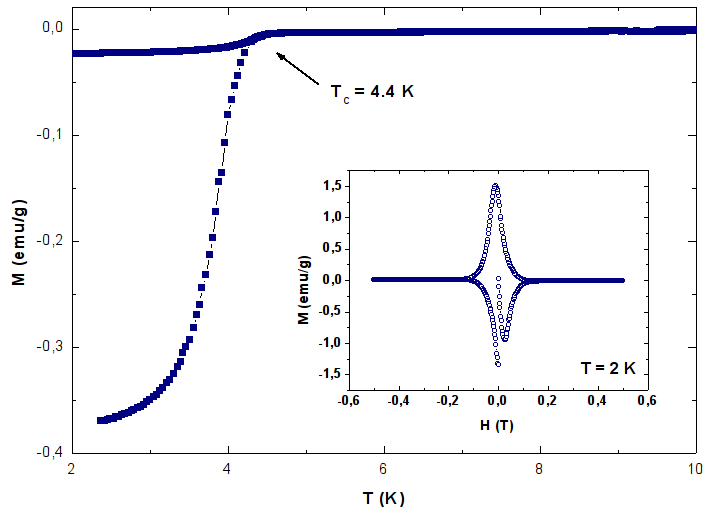}
\caption{\label{fig3} Magnetization as a function of the temperature on a 50 Oe applied field of a polycrystalline Nb$_{2}$BN sample. On the inset isothermal Magnetization as a function of applied field in $T=2$K showing a type II superconducting behaviour.}
\end{figure}

The temperature dependence of the magnetization in zero-field cooled (ZFC) and field cooled (FC) conditions was measured using an applied magnetic field of 50 Oe and is presented in Figure \ref{fig3}. In both ZFC and FC a clear superconducting transition can be seen around 4.4 K. The superconducting volume fraction ($\sim 60$\%) can be estimated within the Meissner state through the dependence of M vs H, since the value of the superconducting state susceptibility (perfect diamagnetism)  is -1/4$\pi$, according to the CGS system. Note that even without considering the demagnetization size susceptibility factor, this result suggests bulk superconductivity. On the FC curve, the flux expulsion of about 7\% indicates a strong flux pinning as expected. Inset of Figure \ref{fig3} displays the isothermal M versus H at $T=2$K that shows a type II superconductivity behaviour.

\begin{figure} [htp]
\includegraphics[width=8.5cm]{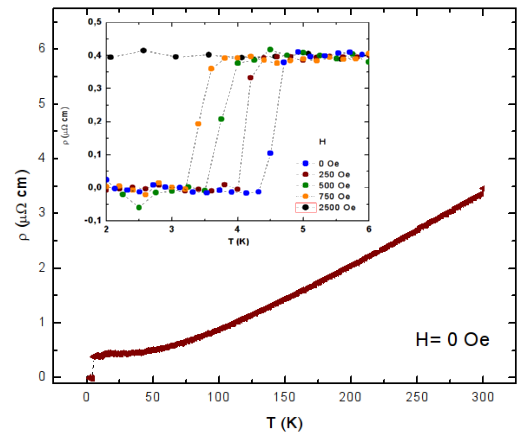}
\caption{\label{fig4} Electrical resistivity ($\rho$) versus temperature of Nb$_{2}$BN from 2 to 300 K displaying the characteristic superconducting transition at 4.4 K. Inset shows the evolution of T$_{c}$ with different applied fields.}
\end{figure}

The typical temperature dependence of the resistivity of Nb$_{2}$BN is shown in Figure \ref{fig4}. A sharp resistivity transition close to 4.4 K (at $\rho=0$, $\Delta T_{c}$ $\sim0.1$ K) is clearly observed at zero applied magnetic field which indicates the high quality of the obtained samples. These results are consistent with those obtained in magnetization measurements. The inset in Fig. \ref{fig4} shows the resistivity dependence on the applied magnetic field. An estimate of the upper critical field at 0 K [$\mu_{0}H_{c2}(0)$] can be made through the Werthamer, Helfand, and Hohenberg (WHH) theory \cite{pool} represented by Eq.~\ref{eq}. According this theory $\mu_{0}H_{c2}(0)$ can be estimated inside the limit of a short electronic mean-free path (dirty limit) and is given by :

\begin{equation}
\label{eq}
\mu_{0}H_{c2}(0)=-0.693T_{c}(dH_{c2}/dT)_{T=T_{c}}
\end{equation}

\begin{figure} [htp]
\includegraphics[width=8.6cm]{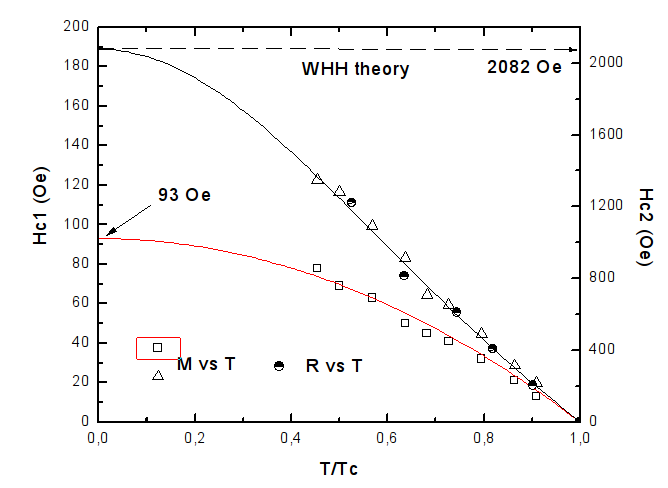}
\caption{\label{fig5} The temperature dependence of $H_{c2}$ and $H_{c1}$ of Nb$_{2}$BN. Experimental points taken from M vs H and R vs T are marked as dots. The red line represents the GL fit for $H_{c1}(T)$ and the black line corresponds to the WHH fit for $H_{c2}(T)$.}
\end{figure}

\begin{figure}
\includegraphics[width=7.2cm]{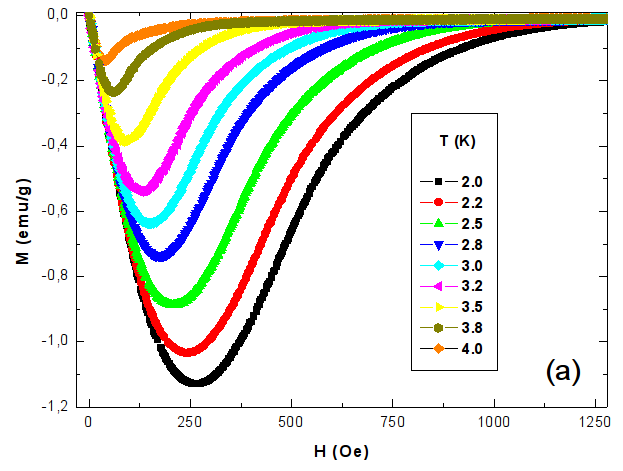}
\includegraphics[width=7.2cm]{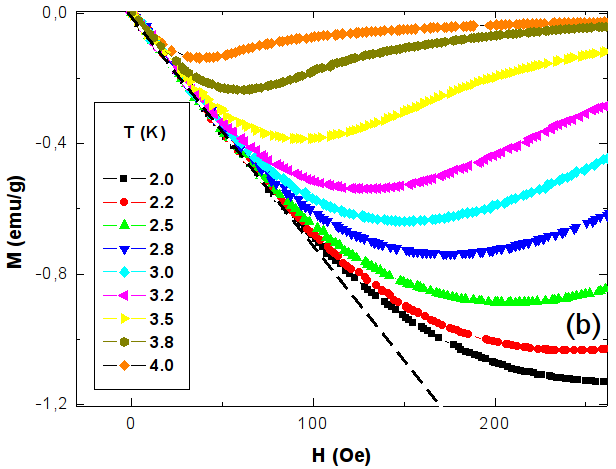}
\caption{\label{fig6} Magnetization versus applied field taken at constant temperatures from 2 to 4 K. The dashed line illustrates the criterion used to determine the difference between M and a the linear behaviour below $H_{c1}$.}
\end{figure}

The temperature dependence of $\mu_{0}H_{c2}(T)$ is shown in Figure \ref{fig5} where the black line represents the conventional fitting of $\mu_{0}H_{c2}(0)$ obtained using the Eq.~\ref{eq}. The critical field $H_{c2}(0)$ is estimated to be $\sim$ 2082 Oe, a value that suggests a type II superconductivity. Another estimate of $H_{c2}(0)$ was obtained by using the data of magnetization versus applied field on different temperatures as shown on Figure \ref{fig6}. The followed criterion consisted of taking the corresponding field of $H_{c2}(T)=0$ for the given temperature, plotting and fitting the values with the equation~\ref{eq} to obtain $H_{c2}(0)$. The result is completely consistent with the resistivity method. The lower critical field (\textit{H$_{c1}(0)$}) was extrapolated from the applied magnetic field dependence of the magnetization at several temperatures shown in Figure \ref{fig6}a, using the Ginzburg-Landau equation \cite{kittel} $H_{c1}(T)=H_{c1}(0)(1-(T/T_{c})^2)$ which gives $H_{c1}(0)\sim$93 Oe. The values of \textit{H$_{c1}$} were determined by examining the divergence from linearity of the slope of the magnetization curve (Fig. \ref{fig6}b), using the criterion \textit{$\Delta$M} = 10$^{-3}$ emu for the difference between the Meissner line and magnetization signal. The calculated values of $H_{c1}(0)$ and $H_{c2}(0)$ allows us to estimate the coherence length and penetration deep through of the Ginzburg-Landau (GL) formulas \cite{kittel}:

\begin{equation}
\label{eq2}
\mu_{0}H_{c1}(0)=\frac{\phi_{0}}{2\pi\lambda_{0}^2}
\end{equation}
\begin{equation}
\label{eq3}
\mu_{0}H_{c2}(0)=\frac{\phi_{0}}{2\pi\xi_{0}^2}
\end{equation}

where $\phi_{0}$ is a quantum flux equal 2.068 x 10$^{-15}$ T.m$^{-2}$, which yields $\xi_{0}$ $\sim$ 425 $\AA$ and $\lambda_{0}$ $\sim 188$ nm  at 0 K. Using the relation $\kappa_{GL}=\lambda_{0}/\xi_{0}$ we obtained the value $\kappa_{GL}= 4.42 $ that agrees with the previous results indicating a type-II superconducting behaviour. 

\begin{figure}
\includegraphics[width=8.5cm]{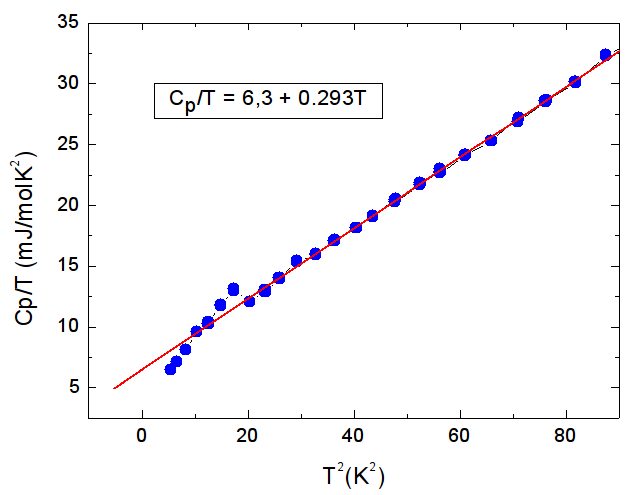}
\caption{\label{fig7}  C$_{p}$/T vs T$^{2}$ curve at zero magnetic field shows the superconducting critical
temperature close to 4.4 K, which is consistent with the electrical transport and
magnetization measurements. The solid red line is the fit of the experimental data to
c = $\gamma$T + $\beta$T$^{3}$ between 2 and 30.0 K.
}
\end{figure}

While the magnetization and resistivity suggest bulk superconductivity in Nb$_{2}$BN,
an anomaly in the specific heat measurement is necessary for confirmation. Figure \ref{fig7}
displays the specific heat divided by temperature (C/T) versus T$^{2}$ at zero applied
magnetic field. A jump in the specific heat is clearly observed on the midpoint of the
transition at Tc = 4.4 K and a small transition length of only $\Delta$T$_{c}\sim$  0.3 K. The consistency
between the magnetization, resistivity, and heat capacity transitions is a clear evidence
of bulk superconductivity in Nb$_{2}$BN. The normal state specific heat (C$_{n}$) is assumed to
have contributions from the standard linear electronic contribution ($\gamma$T) term  and the
cubic phonon ($\beta$T$^{3}$) term.

 In Figure \ref{fig7} the normal state specific heat (C$_{n}$) can be fitted to
the expression $C_{n}=\gamma T+ \beta T^{3}$ by a least squares analysis, with resultant values of $\gamma=6.3$
mJ/mol K$^{2}$ and $\beta =0.293$ mJ/mol K$^{4}$. The $\beta$ value corresponds to a Debye temperature of
$\Theta _{D} \sim 298$ K. The Sommerfeld coefficient $\gamma$ suggests a moderated density of states at the
Fermi level typical of other transition-metal superconductors. 

\begin{figure}[h]
\includegraphics[width=8.5cm]{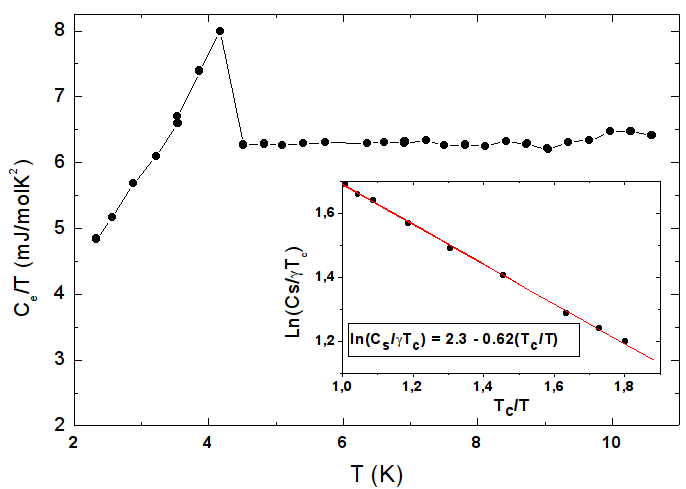}
\caption{\label{fig8} Temperature dependence of the electronic specific-heat divided by
the temperature (C$_{e}$/T). The inset shows the linear fitting of the ln(C$_{s}$/$\gamma$T$_{C}$) against T$_{C}$/T.
}
\end{figure}

A subtraction of the phonon
contribution from the total specific heat allows the analysis of just the electronic
contribution (C$_{e}$), displayed here as C$_{e}$/T versus T in Figure \ref{fig8}. The analysis of the
specific heat anomaly shows the magnitude of the jump at T$_{c}$ to be $\Delta C_{e}/T_{c} \sim 0.88$,
which is significantly smaller than the weak-coupling BCS limit of 1.43. However, this
apparent inconsistency is related with the superconducting fraction estimated
from magnetization measurement displayed in Fig. \ref{fig3}. 
Although the reason for the estimated value from specific heat measurement diverges from BCS prediction is not obvious, the size of the jump strongly suggests that the superconducting behaviour is in a BCS weak coupling limit. Indeed, the exponential behaviour of the electronic component on the superconducting state is in good agreement with the BCS theory as shown in inset of
Figure \ref{fig8}. The linear behaviour of the logarithmic scale versus T$_{c}$/T is totally consistent
with BCS prediction of the superconducting state below $T_{c}$. The comparison with BCS formula for C$_{e}$ below T$_{c}$:

\begin{equation}
\label{eq3}
\frac{C_{e}}{\gamma T_{c}}=8.5 e^{\frac{0.82\Delta (0)}{K_{B}T}}
\end{equation}

which yields an energy gap ($\Delta_{0}$) of 0.65 meV for T$\rightarrow$ 0 and $2\Delta_{0}/K_{B}T_{c}=3.45$ which again is a signal of a weak BCS coupling value. Then, all results from specific heat suggest that Nb$_{2}$BN is a conventional BCS superconducting material. In all BCS superconductors the Cooper-pairing is phonon mediated and the dimensionless electron-phonon constant $\lambda$ can be determined by the McMillan equation \cite{millan} :

\begin{equation}
\label{eq4}
T_{c}=\frac{\theta_{D}}{1.45} exp\lbrace\frac{-1.04(1+\lambda)}{\lambda - \mu^{*}(1+0.62\lambda)}\rbrace
\end{equation}
 
If the Coulomb coupling constant $\mu^{*}$ is considered to be 0.13, which is a usual value, and the Debye temperature from specific-heat is considered 298 K, the value $\lambda\sim0.62$ the obtained value is again in excellent agreement with a weak coupling BCS value for other conventional superconducting materials \cite{other}.   

\begin{center}
\textbf{IV. CONCLUSIONS}
\end{center}

In this paper the superconducting properties of Nb$_{2}$BN were explored. Conventional bulk superconductivity was found at T$_{c}=4.4$ K with superconducting critical fields $H_{c1}(0)=93$ and $H_{c2}(0)=2082$ Oe. Results suggest a conventional type II superconductivity with $\kappa_{GL}= 4.42$ and bulk superconductivity was confirm trough specific heat measurements, showing the values of $\gamma=6.3$
mJ/mol K$^{2}$ and $\beta =0.293$ mJ/mol K$^{4}$ in good agreement with the BCS Theory.  

\begin{center}
\textbf{ACKNOWLEDGEMENTS}
\end{center}

This work was financied by the Brazilian agency CNPq 302892/2011-7, 140804/2012-9 and FAPESP 2010/11770-3

\end{document}